\documentclass[fleqn,usenatbib]{mnras}

\usepackage{newtxtext,newtxmath}

\usepackage[T1]{fontenc}

\DeclareRobustCommand{\VAN}[3]{#2}
\let\VANthebibliography\thebibliography
\def\thebibliography{\DeclareRobustCommand{\VAN}[3]{##3}\VANthebibliography}

\usepackage{graphicx}	
\usepackage{amsmath}	

\newcommand\kms{\rm km\,\rm s^{-1}}
\newcommand\fesc{f_{\rm{esc}}(\rm{LyC})}
\newcommand\xiion{\xi_{\rm{ion},\rm{LyC}}}
\newcommand\hii{H{~\small II}}

\title[A $z=6.8$ galaxy in a self-ionised bubble]{Double-peaked Lyman-$\alpha$ emission at $z=6.803$: a reionisation-era galaxy self-ionising its local H{~\Large II}\, bubble}

\author[R. A. Meyer et al.]{
Romain A. Meyer,$^{1}$\thanks{E-mail: r.meyer.17@ucl.ac.uk}
Nicolas Laporte,$^{2,3}$
Richard S. Ellis, $^{1}$
Anne Verhamme $^{4,5}$ \newauthor
and Thibault Garel $^{4,5}$
\\
$^{1}$Department of Physics and Astronomy, University College London, Gower Street, London WC1E 6BT, UK\\
$^{2}$Kavli Institute for Cosmology, University of Cambridge,Madingley Road, Cambridge CB3 0HA, UK\\
$^{3}$Cavendish Laboratory, University of Cambridge, 19 JJ Thomson Avenue, Cambridge CB3 0HE, UK \\
$^{4}$ Observatoire de Geneve, Universite de Geneve, 51 Ch. des Maillettes, 1290 Versoix, Switzerland \\
$^{5}$ Univ. Lyon1, ENS de Lyon, CNRS, Centre de Recherche Astrophysique de Lyon UMR5574, F-69230, Saint-Genis-Laval, France
}
\date{Accepted 2020 October 12. Received 2020 October 12; in original form 2020 May 14.}

\pubyear{2015}

\begin{document}
\label{firstpage}
\pagerange{\pageref{firstpage}--\pageref{lastpage}}
\maketitle

\begin{abstract}
We report the discovery of a double-peaked Lyman-alpha profile in a galaxy at $z=6.803$, A370p\_z1, in the parallel Frontier Field of Abell 370. The velocity separation between the blue and red peaks of the Lyman-$\alpha$ profile ($\Delta v=101_{-19}^{+38} (\pm48)\,\kms$) suggests an extremely high escape fraction of ionising photons $> 59(51)\% (2\sigma)$. The spectral energy distribution indicates a young ($50$ Myr), star-forming ($12\pm 6 \, \rm{M}_\odot \rm{yr}^{-1}$) galaxy with an IRAC excess implying strong [OIII]+H$\beta$ emission. On the basis of the high escape fraction measured, we demonstrate that A370p\_z1 was solely capable of creating an ionised bubble sufficiently large to account for the blue component of its Lyman-alpha profile. We discuss whether A370p\_z1 may be representative of a larger population of luminous $z\simeq$7 double-peaked Lyman-alpha emitting sources with high escape fractions that self-ionised their surroundings without contributions from associated UV-fainter sources.
\end{abstract}




\section{Introduction}
\label{sec:intro}
Cosmic reionisation marks the last phase transition of the Universe when the intergalactic medium (IGM) was reionised, thus ending the so-called Dark Ages. The timing of reionisation is now constrained to $5.5\lesssim z \lesssim 15$ by a variety of probes \citep{Stark2010,Becker2015b, PlanckCollaboration2018, Banados2019}. Yet the sources capable of emitting sufficient ionising photons by $z\sim 5.5$ continue to be the subject of debate \citep[e.g. see Section 7\&8 of ][for a review]{Dayal2018}. A widely-held view is that intrinsically UV-faint galaxies are the primary contributors, typically leaking $\sim 10\%$ of their Lyman continuum (LyC) photons to the intergalactic medium \citep[ e.g.][]{Robertson2015, Finkelstein2019, Dayal2020}. However, to match the relative rapid decline of the neutral fraction at late times, rarer, luminous sources may play a significant role \citep[e.g][]{Naidu2020}. The issue remains unsolved as there is yet no direct way of measuring the escape fraction, $\fesc$, of LyC radiation at high redshift.

At $z<4$, LyC leakers are being studied in detail providing new insight into the physical conditions under which ionising photons can escape. A picture is emerging where LyC leakage may be linked to the [OIII]/[OII] emission line ratio \citep[][but see also \citet{Bassett2019}]{Nakajima2018,Nakajima2020}, varies geometrically due to low-column density channels which allow the photons to escape \citep{Fletcher2019} and is correlated with the Lyman-$\alpha$ emission line profile \citep[e.g.][]{Verhamme2015, Izotov2018}. Of particular interest is the correlation with the velocity separation in double-peaked Lyman-$\alpha$ profiles \citep[e.g.][]{Izotov2018}. As Lyman-$\alpha$ photons are scattered and Doppler-shifted in dense neutral gas before emerging out of resonance on either the blue or red side of the peak, the double peak separation is linked to the H{~\small I} column density that controls the LyC escape fraction \citep{Verhamme2015,Kakiichi2019}. Moreover, after the Lyman-$\alpha$ photons escape, only a modest amount of neutral gas in the IGM would absorb the blue wing \citep{Dijkstra2014}. Double-peaked Lyman-$\alpha$ emitters thus also constrain the size of any associated ionised bubble \citep[e.g.][]{Mason2020}.

Thus far, only two galaxies at $z>6$ (NEPLA4, $z=6.54$ \citep{Songaila2018} and COLA1, z=$6.59$ \citep{Hu2016,Matthee2018}) are known to have a double-peaked Lyman-$\alpha$ profile. \citet{Bosman2020} also recently reported a double-peaked profile in a $z\sim5.8$ Lyman-Break selected galaxy, Aerith B, in the near-zone of a quasar. The peak velocity separation measured in the three Lyman-$\alpha$ profiles have provided useful estimates of $\fesc$ in high-redshift galaxies. 
In this paper, we report the discovery of a new galaxy presenting a double-peaked  Lyman-$\alpha$ profile at $z=6.803$, deeper in the reionisation era than those above. Its Lyman-$\alpha$ profile indicates a much larger escape fraction and a capability to self-ionise its local \hii\, bubble. We discuss whether it is representative of those sources that ended cosmic reionisation. Throughout this paper, magnitudes are in the AB system \citep{Oke1974}, and we use a concordance cosmology with $H_0 = 70, \Omega_M = 0.3, \Omega_\Lambda = 0.7$. We refer to proper (comoving) kiloparsecs and Megaparsecs as p(c)kpc and p(c)Mpc.

\section{Observations}
\label{sec:obs}
The target of this study was originally observed as part of a search for rest-frame UV lines signalling AGN activity in bright $z\sim7$ galaxies (X-Shooter/VLT, ID: 0100.A-0664(A), PI: Laporte). Following earlier detection of He{~\small II} emission in a galaxy with evidence for strong [OIII] and H$\beta$ emission lines \citep{Laporte_2017b}, we searched for similar sources using data from the Hubble and Spitzer Space Telescopes
in the Frontier Fields survey \citep{Lotz2017}, applying selection criteria defined in \citet{Bouwens2015}. Possible evidence for intense [O III] and $H\beta$ line emission was considered via excess emission in the appropriate IRAC bandpasses \citep[see][]{Labbe2013, Smit2014}. 

Spectroscopic follow-up was conducted with both X-Shooter/VLT and ALMA (\citealt{Laporte_2017b}, \citealt{Hashimoto2018a}, \citealt{Laporte2019}) to determine the redshifts, star-formation rates (SFR) and other properties. Among this sample, one bright galaxy (F125W$=25.16$, $z_{\rm{phot}}$=7.14$\pm0.8$), hereafter A370p\_z1, was observed with X-Shooter/VLT in service mode in October 2018. Observing blocks were defined in order to maximise the exposure time in the NIR arm (t$_{\rm NIR}=900$s, t$_{\rm VIS}=819$s and t$_{\rm UVB}=756$s). The target was centred in a $0.9''$ slit using a blind offset from a nearby bright star. After discarding time in poor seeing, the usable exposure time in the VIS arm was $6.3$hrs. 

\begin{figure}
    \centering
    \includegraphics[width = 0.5\textwidth]{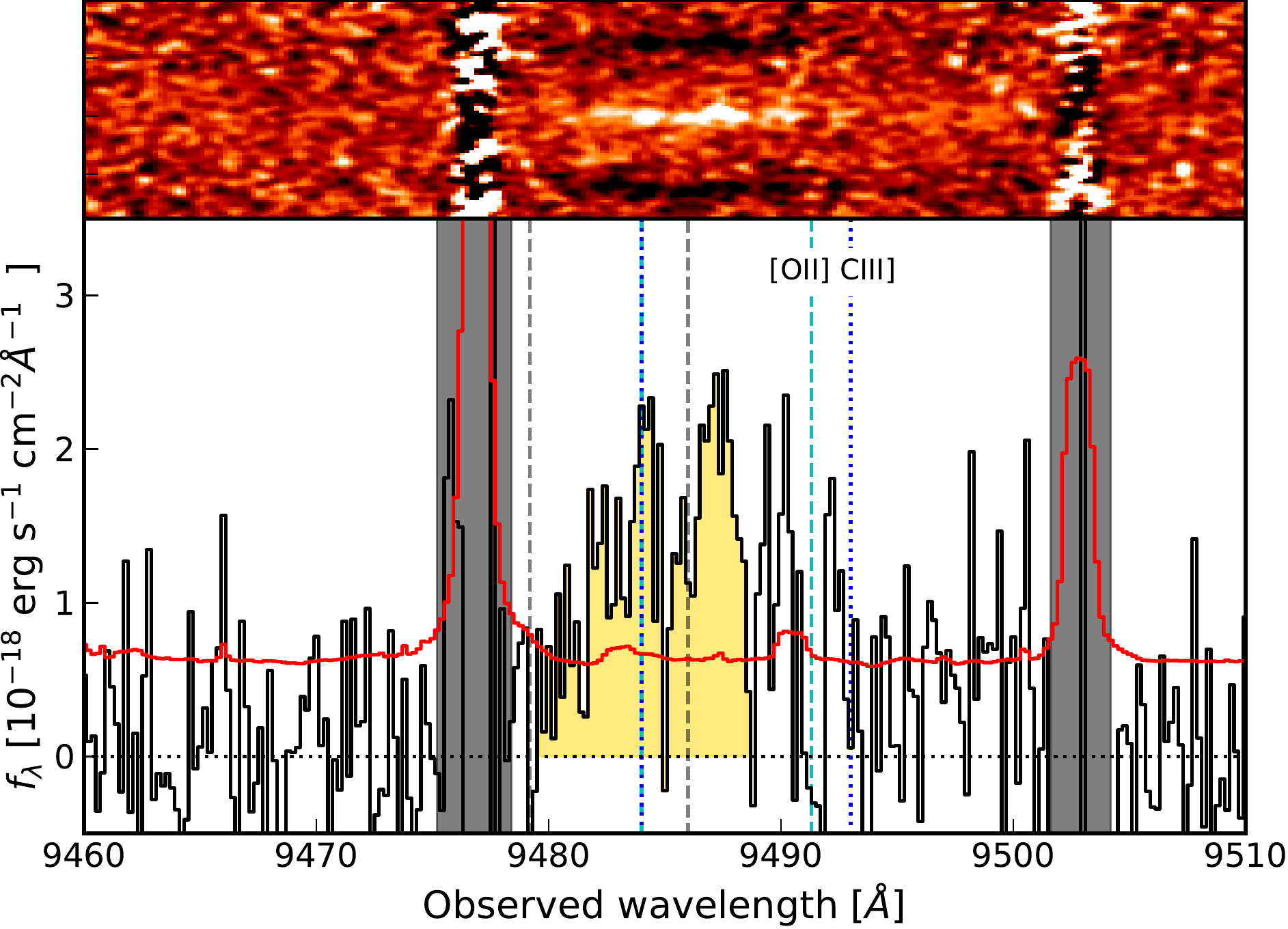}
    \caption{2D spectra of A370p\_z1, showing double-peaked Lyman-$\alpha$ emission at $z=6.803$ with two negative counterparts arising from the telescope dither pattern. The lower panel shows the 1D spectrum (black) and error array (red) with OH sky lines masked in grey. The two peaks are  highlighted in yellow. Vertical dotted lines show the maximum extent of the blue wing and the mid-point of the two peaks. The velocity separation of emission line doublets of potential low-redshift redshift interlopers is illustrated by cyan dashed ([OII]$\lambda\lambda 3727,3729$) and blue dotted (CIII]$\lambda\lambda 1907,1909$) vertical lines, with the first peak of the doublets placed at the observed blue peak wavelength $\lambda=9484 $ \AA.}
    \label{fig:observed}
\end{figure}

The spectroscopic data were reduced using standard X-Shooter \textsc{ESOReflex} recipes (v3.3.5). Flux calibrated 2D spectra were stacked using \textsc{IRAF}'s \textit{imcombine} and visually inspected for emission lines by two authors (RAM, NL). Stacking with custom \textsc{ESOReflex} and \textsc{Python} recipes produced similar results. The stacked 2D spectrum was optimally extracted \citep{Horne1986} with a boxcar aperture of 1.6'' (10 pixels) revealing an emission line doublet at $9484,9487$ \AA \, (Fig. \ref{fig:observed}). No other line was found in the X-Shooter data.

\begin{figure}
    \includegraphics[width = 0.5\textwidth, trim={3.2cm 0cm 0cm 0cm},clip]{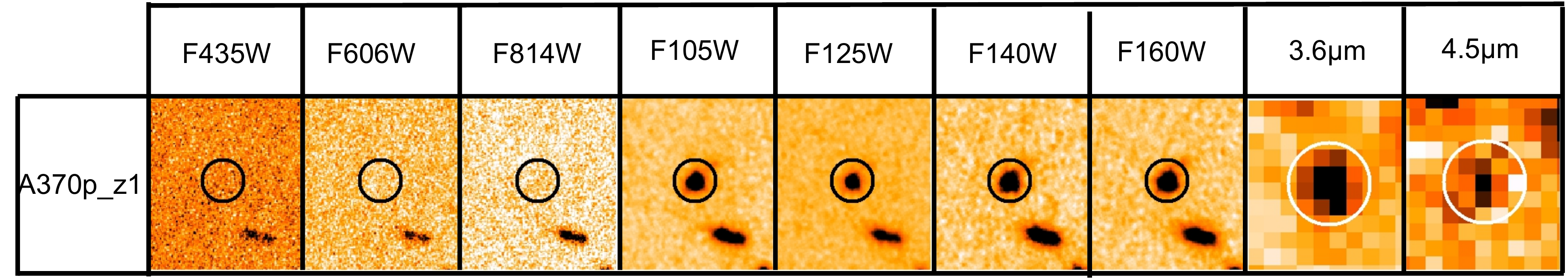}\\ 
     \includegraphics[width = 0.5\textwidth, trim={1cm 1cm 1cm 2cm},clip]{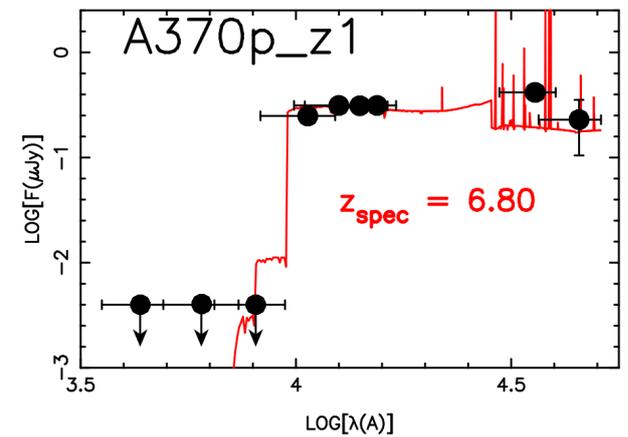}
    \caption{\textbf{Top panel:} Frontier Field image stamps (HST + Spitzer, $\sim 3.5\times 3.5$ arcsec$^2$) of A370p\_z1, showing a clear drop in F105W-F814W, typical of $z\gtrsim 6.5$ galaxies. The Spitzer channels have been decontaminated from the contribution of a southern object (see Sec. \ref{sec:phys_nature}).  \textbf{Lower panel:} Spectral energy distribution based on the photometry (black) with a BAGPIPES fit (red) adopting a redshift $z=6.80$ from the Lyman-$\alpha$ profile. Note a 3.6$\mu \rm m$ excess likely due to [OIII]+H$\beta$ emission, claimed to be an indicator of high $\fesc$ \citep[e.g.][]{Faisst2016,Izotov2018,Tang2019,Nakajima2020}}
    \label{fig:photometry}
\end{figure}
The trough between the two peaks is about twice the X-Shooter resolution for the adopted $0.9''$ slit (34 km s$^{-1}$). To verify that the inter-peak absorption is significant, we compute the residuals of the dip pixels with respect to the flux of the smaller peak (the blue peak). The $\chi^2$ statistic gives a only $P(\chi^{2}) = 0.00013$ probability that the dip is consistent with Gaussian residuals around the blue peak maximum. The inter-peak absorption is therefore significant at $3.8\sigma$. However, this statistic does not guarantee that the double-peaked profile would be selected by eye when inspecting the 2D spectra. In order to recognise a double-peak, observers look for a few significantly absorbed pixels, preferably consecutive, in-between the peaks. We resampled the spectrum between $\lambda=9480,9489$\AA\, assuming a Gaussian noise distribution with variance drawn from the error array. We then identified the maximum pixels on either side of $\lambda = 9485$\AA\, to find the profile peaks. Counting how many pixels are $>2\sigma$ below the average of the peaks' maximum flux, we found that in $\sim 95\%$ ($2\sigma$) of the resampled spectra, there are at least four pixels satisfying this criteria, and at least two are contiguous. We used this bootstrap resampling technique to obtain robust errors on the peak velocity separation $\Delta v = 101_{-19}^{+38} \,\kms$. We note that these errors might be slightly underestimated because the peak separation is measured from the maximum of the peaks. A more conservative error estimate based on the resolution of the spectra ($33.5\, \kms$) gives $\Delta v = 101 \pm 48 \,\kms$.

The observed peak separation rules out a $z\sim1.54 $ [OII] $\lambda\lambda \, 3727,3729\,$\AA \,interloper ($\Delta v = 225\,\kms$) and a $z\sim 4.15$ C{~\small III}]$\lambda\lambda\, 1907, 1909\,$\AA\,  doublet ($\Delta v = 314\,\kms$). A low-redshift interpretation would predict other lines in the UV, VIS and NIR arms but none was found (e.g. [OIII] and [O II] are detectable with X-Shooter up to $z\sim3.8$ and 5.4, respectively). A high redshift solution is also consistent with the Lyman break seen in the SED (Fig. \ref{fig:photometry}); a dusty source with a Balmer break at $z\sim1.5-2$ is inconsistent with the flat SED redwards of $1.5\mu$m. Although it is possible the peaks come from different locations in a single galaxy or a merger, our 2D spectral data indicates both peaks are co-spatial. 
\begin{figure}
    \centering
    \includegraphics[width = 0.48\textwidth]{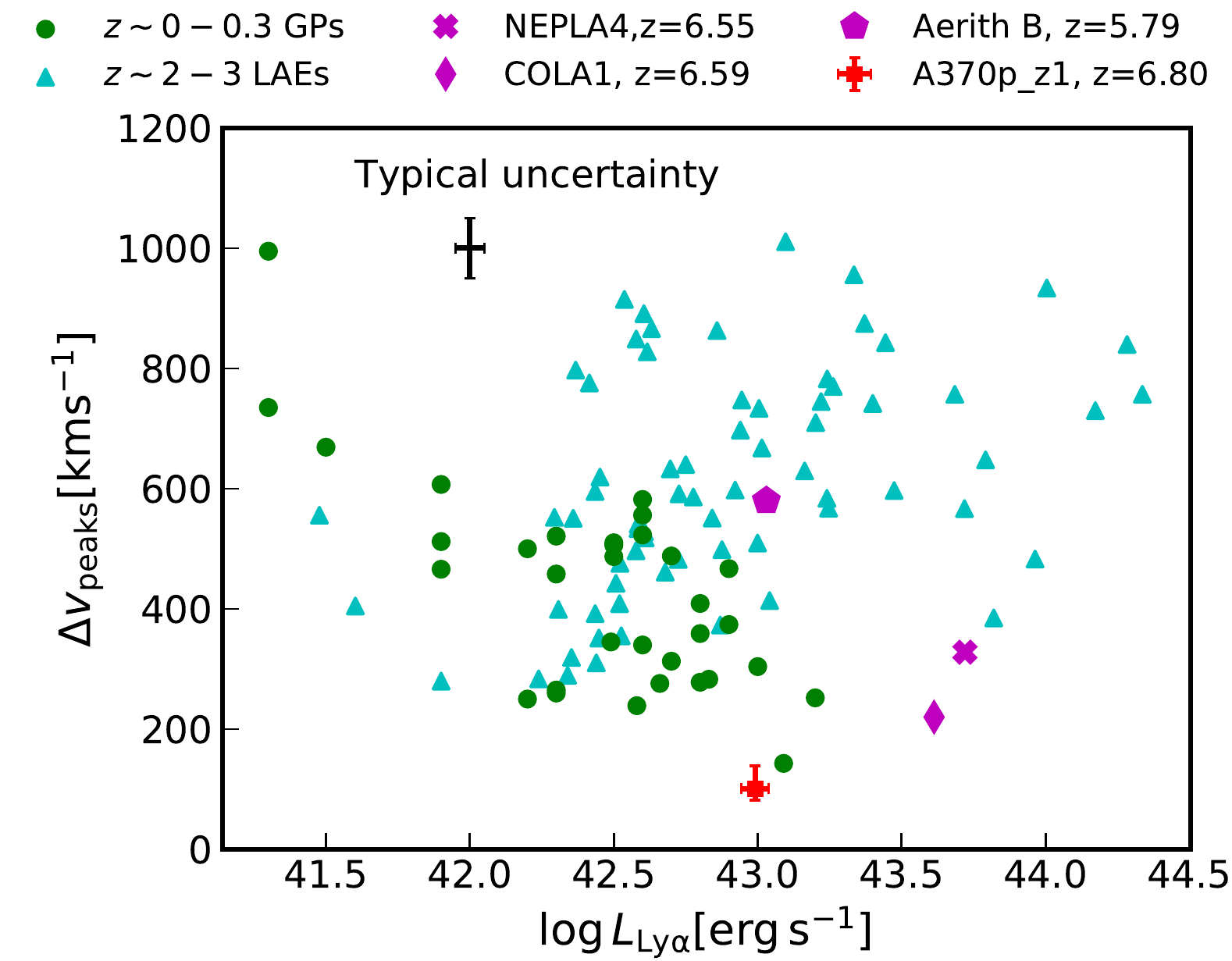}
    \caption{Lyman-$\alpha$ peak separation versus Lyman-$\alpha$ luminosities for $z\sim 0$ Green Peas  \citep[green circles, ][]{Yamada2012} and $\sim 2-3$ double-peaked LAEs \citep[cyan triangles,][]{Yang2017,Kulas2012,Hashimoto2015}, Aerith B \citep[magenta pentagon, ][]{Bosman2020}, NEPLA4 \citep[magenta cross, ][]{Songaila2018}, COLA1 \citep[magenta diamond][]{Matthee2018} and A370p\_z1 (red square). }
    \label{fig:lowz_comparison}
\end{figure}

We therefore conclude that A370p\_z1 has a double-peaked Lyman-$\alpha$ profile at $z$=6.803 (taken as the mid-point of the two peaks\footnote{\citet{Verhamme2018} show that for all double-peaked profiles with known systemic redshift reported in the literature, the systemic redshift always falls close to the mid-point of the peaks, in agreement with the findings of radiative-transfer simulations.}) with a peak velocity separation of $\Delta v = 101_{-19}^{+38} \,\kms$. The Lyman-$\alpha$ rest-frame luminosity ($(9.8 \pm  1.0) \times 10^{42} \rm{erg\, s} ^{-1}$) and equivalent width ($\rm EW_{Ly\alpha}= 43\pm 4$ \AA) are similar to those seen in $z\sim 0$ Green Peas  \citep{Yang2017,Izotov2018} or $\sim 2-3$ double-peaked Lyman-$\alpha$ emitters (LAEs) \citep[e.g. ][see Fig. \ref{fig:lowz_comparison}]{Yamada2012,Kulas2012,Hashimoto2015}. Finally, \citet{Matthee2018} raised the possibility that high-redshift double-peaked Lyman-$\alpha$ could be potentially caused by a  foreground absorber in a standard (red-wing only) Lyman-$\alpha$ line. However, the skewness of the red and (blue) peak is $S=0.70\pm 0.24 \,(-0.32\pm0.23)$ which is higher than the $S>0.15$ threshold used for LAEs \citep{Kashikawa2006}. The skewness of the peaks also disfavours the merger interpretation. We searched for evidence of a hard ionisation spectrum or AGN activity but, at the expected location of N{~\small IV} 1240 \AA, C{~\small IV} 1549 \AA, He{~\small II} 1640 \AA, C{~\small III}]$\lambda\lambda 1907, 1909$ \AA\, we do not find any significant emission lines (Fig. \ref{fig:observed_extralines}). Table \ref{table:properties} summarises the properties of A370p\_z1. The uncertainties are derived using the spectral resolution ($R\sim 8900$) and the error array, except for the peak velocity separation which comes from bootstrapping. 

\begin{table}
    \centering    
    \caption{Properties of A370p\_z1. Limits are quoted at the $2\sigma$ level. The peak velocity separation and associated escape fraction are given with bootstrap errors and resolution errors (see Section \ref{sec:obs}).}
    \begin{tabular}{lr} \hline
    RA & 02h40m14.1s   \\
    DEC & -01d37m14.3s \\ \hline
   \multicolumn{2}{l}{Emission lines}  \\ \hline
     $F_{\rm Ly\alpha}$ & $(18.4\pm 1.9)\times 10^{-18} \rm{\,erg\,cm}^{-2}\rm{s}^{-1}$\\
     $F_{\rm NV}$& $< 1.9\times 10^{-18}\rm{\,erg\,cm}^{-2}\rm{s}^{-1} $\\
     $F_{\rm CIV}$& $< 1.3\times 10^{-18}\rm{\,erg\,cm}^{-2}\rm{s}^{-1} $\\
     $F_{\rm HeII}$& $< 2.6\times 10^{-18}\rm{\,erg\,cm}^{-2}\rm{s}^{-1}$\\
     $F_{\rm CIII}$& $< 0.7\times 10^{-18}\rm{\,erg\,cm}^{-2}\rm{s}^{-1}$ \\ \hline
    \multicolumn{2}{l}{Lyman-alpha profile } \\ \hline
        $z_{Ly\alpha}$ & $6.803$    \\
        $\Delta v_{\rm Ly\alpha}$   & $101_{-19}^{+38} (\pm48)\,\kms$.\\ 
        FWHM$_{\rm blue}$  & $82\pm 48 \, \kms$ \\
        FWHM$_{\rm red}$  &$120\pm 48 \, \kms$\\
        f$_{\rm blue}$ &   $(7.4\pm 1.9)\times 10^{-18} \rm{\,erg\,cm}^{-2}\rm{s}^{-1}$ \\
        f$_{\rm red}$ & $(10.8\pm 2.4)\times 10^{-18}\rm{\,erg\,cm}^{-2}\rm{s}^{-1}$ \\
        Blue/red flux ratio & $0.69 \pm 0.24$  \\
        Blue peak skewness & $-0.32\pm0.23$ \\
        Red peak skewness & $0.70\pm0.24$\\
        $\rm L_{Ly\alpha}$ (rest-frame) & $(9.8 \pm  1.0) \times 10^{42} \rm{erg\, s} ^{-1}$ \\
        $\rm EW_{Ly\alpha}$ (rest-frame) & $43\pm 4 \,$ \AA \\
        $\fesc$ \citep{Izotov2018} & $> 0.59(>0.51)$ \\  
        $\fesc$ (RASCAS) & $0.99$ \\\hline
   \multicolumn{2}{l}{Photometry and SED fitting (BAGPIPES)} \\ \hline
        $m_{\rm F435W}$ & $<29.90$ \\
        $m_{\rm F606W}$ & $<29.80$  \\
        $m_{\rm F814W}$ & $<30.00$ \\
        $m_{\rm F105W}$ & $25.41 \pm 0.01$ \\
        $m_{\rm F125W}$ & $25.16 \pm 0.01$ \\
        $m_{\rm F140W}$ & $25.17 \pm 0.01$ \\
        $m_{\rm F160W}$ & $25.16 \pm 0.01$ \\
        $m_{3.6\mu}$ & $24.85\pm0.14$ \\
        $m_{4.5\mu}$ & $26.19\pm0.50$ \\
        $M_{\rm UV}(m_{\rm{F105W}})$ & $-21.5 \pm 0.1$  \\
        $M_{*}$ & $(6.55^{+0.14}_{-0.10}) \times 10^{9} \rm{M}_\odot$ \\
        SFR & $12 \pm 6 \, \rm{M}_\odot\,\rm yr^{-1}$ \\
        Age & $50\pm4 \, \rm Myr $ \\ \hline
    \end{tabular}
    \label{table:properties}
\end{table}

\section{Results}
\subsection{The nature of  A370p\_z1}
\label{sec:phys_nature}
We first characterised A370p\_z1 by utilising the available deep \textit{HST} and \textit{Spitzer} Frontier Fields imaging (Fig. \ref{fig:photometry}, upper panel). We extracted the spectral energy distribution (SED) following the method described in \citet{Finkelstein2013}. We neglect any lensing magnification as A370p\_z1 is in a parallel field and thus far from the cluster Abell 370. The \textit{Spitzer} 3.6$\mu$m and 4.5$\mu$m images are contaminated by a point source 1.5'' to the south-east. We used GALFIT \citep{Peng2010} to remove its contaminating contribution and applied a standard aperture correction. We fit the SED using BAGPIPES \citep{Carnall2018}, experimenting with several star formation histories (SFH) adopting single (constant, exponential, burst) and two component models (constant + burst ; exponential + burst). The best-fit SED was a constant SFH model with the following properties : age = $(50\pm4)\, \rm{Myr}$, M$_{\star}=(6.55^{+0.14}_{-0.10})\times$10$^{9}\, \rm{M}_{\odot}$ and a SFR $= (12\pm6) \,\rm{M}_{\odot}\rm{yr}^{-1}$ (Fig. \ref{fig:photometry}, lower panel, and Table \ref{table:properties}). We found no preference for an exponentially-declining SFH or a single-burst model. The flux limits  at the rest-frame UV lines  positions are consistent with the line fluxes from the best-fit SED (see further Table \ref{table:properties}).

As discussed in the Introduction, a small separation for a double-peaked Lyman-$\alpha$ profile is a strong indicator of a high LyC escape fraction in low-redshift analogues \citep[e.g][]{Gronke2017,Verhamme2017}. However, the tight empirical relation found by \citet{Izotov2018}, 
\begin{equation}
    \fesc = \frac{3.23\times10^{4}}{\Delta v_{Ly\alpha}^2} +\frac{1.05\times10^{2}}{\Delta v_{Ly\alpha}} + 0.095 \text{      ,}
\end{equation}
may not apply for the range $v_{\rm peaks} \lesssim 150\, \kms$ which was not probed by their observations and where their relation would predict an unphysical $\fesc >100\%$. We therefore put a maximum of $100\%$ to the polynomial function so it does not result in unphysical values. We then compute the escape fraction for each of the resampled spectra (see Section \ref{sec:obs}) to obtain a $2\sigma$ lower limit on A370p\_z1 LyC escape fraction $\fesc >$ 59\%. Using the conservative error from the X-Shooter resolution gives $\fesc>51\%$ ($2\sigma$).

\begin{table}
    \centering
    \caption{Shell model parameter grid searched.}
    \begin{tabular}{l|r}
        \hline 
        $b\, [\kms]$ & $20,80,140$ \\
        $v_{\rm{exp}} \, [\kms]$ & $0,20,50$ \\
        $\log \rm{N_{HI}/[cm^{-2}]}$ & $15,16,17,18 $ \\
        $\tau_{\rm{d}}$ & $0,0.5,1$ \\
        FWHM$_{Ly\alpha}\, [\kms]$ & $100,200,300,400,500$\\ \hline
    \end{tabular}
    \label{table:param_shell}
\end{table}

In order to better estimate the escape fraction, we compare the observed profile with double-peak shell models.
We use the RASCAS 3D Monte-Carlo code \citep{Michel-Dansac2020} to generate a grid of Lyman-alpha radiation transfer simulations in spherical geometries, allowing for static and expanding gas configurations. In these typical shell models \citep[e.g.][]{Dijkstra2006,Verhamme2008}, H{~\small I} gas and dust are distributed homogeneously around a central point source. The shell is described by four physical parameters, namely the expanding velocity $v_{\rm{exp}}$, the HI column density $N_{HI}$, the dust opacity $\tau_d$, and the Doppler parameter $b$ which accounts for the thermal/turbulent gas motions (see Table \ref{table:param_shell} for the parameter grid used). The intrinsic emission is assumed to be a Gaussian line centred on the systemic redshift with a width set by the FWHM. Given the nearly symmetric double peak profile of A370p\_z1 and the small peak separation, we restrict our analysis to relatively small $N_{HI}$ and $v_{\rm{exp}}$ values because it is well-known that high column densities and shell velocities would significantly broaden the line and erase the blue peak respectively \citep{Verhamme2006}. We perform a quantitative comparison between the observed line profile and the models using the $\chi^2$ statistic. We find that models minimising the reduced $\chi^2$ preferentially select low $N_{HI}$ ($\log \rm{N_{HI}/[cm^{-2}]}=15$), static geometries ($v_{exp}=0$), low dust content, small b values ($b=20$) and relatively broad input lines ($200 \, \kms< \rm{FWHM}_{Ly\alpha} < 400\, \kms$).

\begin{figure}
    \centering
    \includegraphics[width=0.45\textwidth]{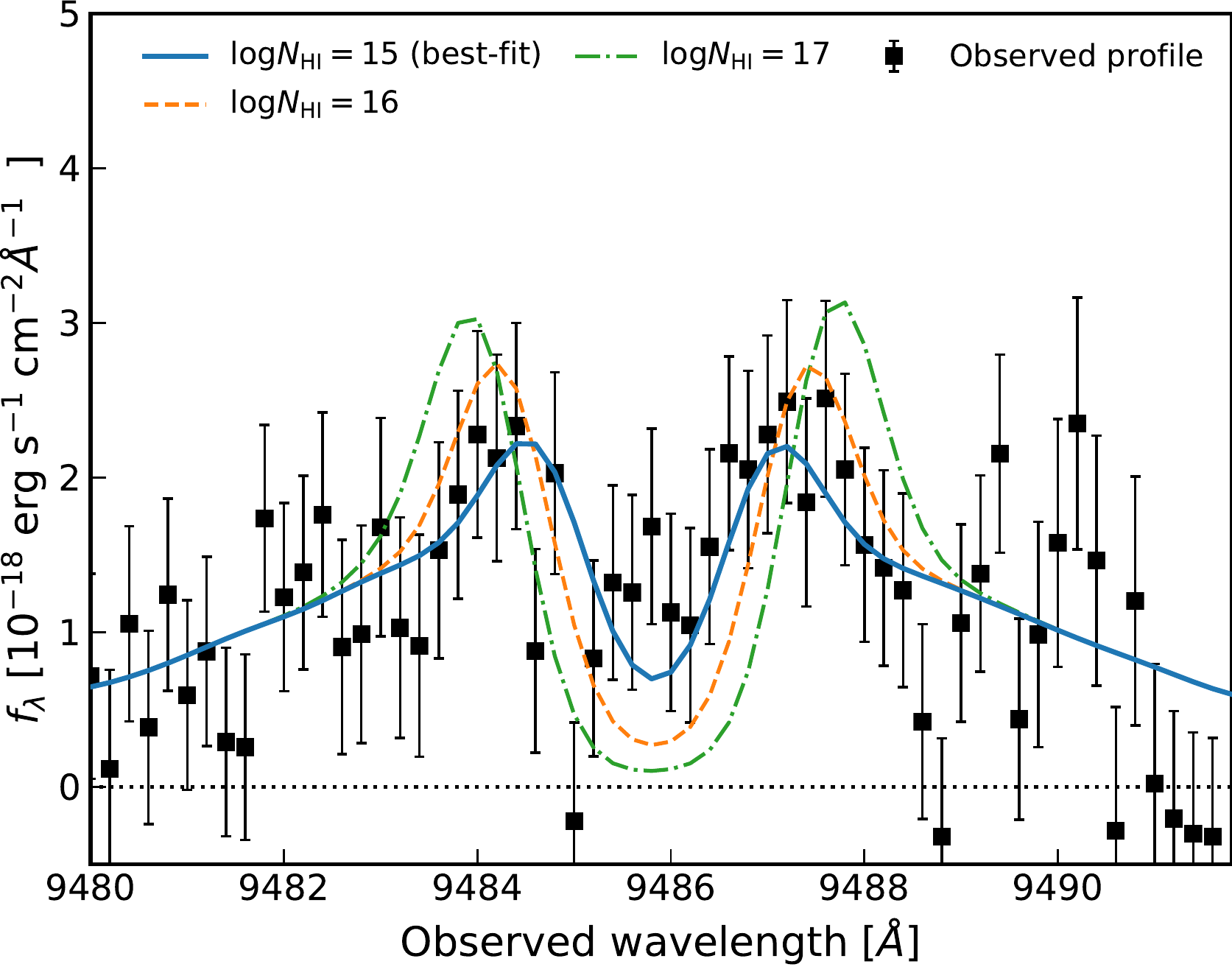}
    \caption{Comparison of RASCAS models, smoothed by the resolution of X-Shooter, of Lyman$\alpha$ transfer through non-expanding shells of homogeneous gas with the observed Lyman-$\alpha$ profile. The best-fit model (blue) has  $\log \rm{N_{HI}/[cm^{-2}]} = 10^{15}$ which correspond to a high LyC escape fractions $\fesc = 99\%$. We also show examples of models with $\log \rm{N_{HI}/[cm^{-2}]} = 10^{16}(10^{17})$ (dashed orange, dashed-dotted green), which would imply $\fesc = 94\%(53\%)$. }
    \label{fig:models}
\end{figure}

We show the best fit model in Figure \ref{fig:models} which corresponds to the following parameter set: $\log \rm{N_{HI}/[cm^{-2}]}=15, b=20\, \kms, v_{exp}=0, \rm{FWHM}=300\,\kms$. 
We can derive the LyC escape fraction from the best-fit column density $\fesc = \exp\left(-\sigma_{912}\rm{N_{HI}}\right)$ = 99\%, where $\sigma_{912}= 6.35\times 10^{-18}$ is the H{~\small I} photoionisation cross-section at the Lyman limit. While several models could match the positions of peak emission within the errors, the $\log \rm{N_{HI}/[cm^{-2}]} = 10^{15}$ model is the only one to reproduce the shallow central depression of the profile which is the important signature of a low H{~\small I} opacity (Figure \ref{fig:models}). Searching a finer parameter grid is beyond the scope of this paper, but we note that even when adopting a higher column density $\log \rm{N_{HI}} = 10^{16}$ cm$^{-2}$, the escape fraction remains very high (94\%). 

\subsection{Did A370p\_z1 self-ionise its local H{~\small  II} bubble?}

The detection of the blue peak in the Lyman-$\alpha$ indicates A370p\_z1 sits in a large ionised bubble, otherwise the damping wing of even a partially neutral IGM would have absorbed it. Given its high escape fraction, we now consider whether A370p\_z1 could have self-ionised its local \hii\, bubble. 

The blue wing extends to $\lambda = 9479.2$ \AA, $\approx 215\, \kms$ from the line centre, corresponding to a physical distance $r_{\rm HII} >0.26 \pm 0.04$ pMpc. This estimate neglects any velocity offset between the Lyman-$\alpha$ absorption dip from which we have derived the redshift and the systemic redshift as defined by more reliable tracers such as nebular absorption lines. These velocities are however found to be $\lesssim 200\, \kms$ \citep{Gazagnes2020}, which, in the worst case, would therefore require a larger ionised bubble for the blue wing to be transmitted ($\lesssim 0.5$ pMpc). We also neglect peculiar velocities of the galaxy with respect to the ionised bubble gas which would redshift(blueshift) the Lyman-$\alpha$ photons and decrease(increase) the bubble size needed for the blue wing to escape.

We now estimate the volume that could have been reionised by A370p\_z1 by redshift $z= 6.803$ and whether its radiation is sufficient to reduce the opacity of the surrounding gas to permit the blue wing of Lyman-$\alpha$ to escape. Assuming no recombination, the ionised bubble (Str\"omgren sphere) created by a single galaxy in the reionisation era is \citep[e.g.][]{Cen2000}
\begin{equation}
    R_S \approx \left( \frac{4 \fesc \xiion L_{UV} t_{\rm{em}}}{3\pi\langle n_{HI}\rangle}\right) ^{1/3}
\end{equation}
where $t_{\rm{em}}$ is the duration of LyC leakage from a source with intrinsic ionising efficiency $\xiion$ and escape fraction $\fesc$ and $\langle n_{HI}(z)\rangle\approx 8.5\times10^{-5} \left( \frac{1+z}{8}\right)^3 \rm{cm^{-3}}$ is the mean hydrogen density of the IGM. The typical ionising efficiency of $M_{\rm UV} = -22$ galaxies at $z\sim 5$ is $\log \xi_{ion}\simeq 25.4$ cgs \citep{Bouwens2015}. However, recently it has been claimed that some $z>7$ galaxies have enhanced ionising efficiencies \citep{Stark2015, Stark2017}. We therefore derive an estimate of the ionising efficiency from the Lyman-$\alpha$ line following \citep{Sobral2019} and find $\log \xi_{ion}\simeq 26.4$ cgs. In the following, we indicate results based on the higher ionising efficiency in parenthesis.
Assuming an $\fesc \approx 0.9$,
$50$ Myr is a sufficient time for A370p\_z1 to create an ionised bubble with radius $R_S \simeq 0.86(1.10)\, \rm{pMpc}$. This is more than three times larger than the distance at which the blue wing of Lyman-$\alpha$ is still transmitted. Therefore, it is plausible that A370p\_z1 is able to self-ionise its surrounding bubble, even if the escape fraction was $\approx90\%$ only for a small fraction of its lifetime (e.g. $R_S(t_{\rm em}= 0.2 t_{\rm{age}}) \approx 0.50(1.05)$ pMpc). 

Being able to grow a Str\"omgren sphere larger than the distance required for the blue wing of Lyman$-\alpha$ to redshift out of resonance is a necessary but not sufficient condition for A370p\_z1 to be solely responsible for its ionised bubble. This is because the Gunn-Peterson optical depth is virtually zero for neutral fractions as low as $10^{-4.5}$ \citep[see][for a review]{Becker2015}. Therefore, in the absence of an elevated photoionisation rate, the blue peak would readily be resonantly absorbed by even small pockets of neutral gas within the ionised bubble. We therefore examine whether A370p\_z1 can maintain such a high photoionisation rate at the edge of its bubble or if an additional population of clustered UV-faint galaxies is required.

Following \citet[][and references therein]{Kakiichi2018}, the local photoionisation rate due to A370p\_z1 is
\begin{eqnarray}
   \Gamma_{\rm HI}^{\rm{A370p\_z1}}(r)  & = & \frac{\alpha_g \sigma_{\rm{912}}}{\alpha_g+3} \frac{\fesc \xi_{\rm{ion}} L_{\rm{UV}}}{4\pi r^2} \rm{e}^{-r/\lambda_{\rm {mfp}}} \nonumber \\
    & \simeq & 0.8(7.4)\times 10^{-11}\left(\frac{r}{0.1 \rm{pMpc}}\right)^{-2}   \text{s}^{-1} 
\end{eqnarray}
where $\alpha_g$ is the extreme UV spectral slope, and $\lambda_{\rm{mfp}}$  is the mean free path of LyC photons. We assume $\alpha_g =2$ \citep[e.g.][]{Kuhlen2012} and $\lambda_{\rm{mfp}}\simeq 6.0 \left( \frac{1+z}{7}\right)^{-5.4} \rm{pMpc}$ \citep{Worseck2014}.
The fluctuating Gunn-Peterson approximation links the photoionisation rate $\Gamma$ to the Lyman-$\alpha$ opacity 
$\tau_\alpha \simeq 11 \Delta_b^{2-0.72(\gamma-1)}  \left( \frac{\Gamma_{\rm HI}}{10^{-12} \text{ s}^{-1}}\right)^{-1}  \left( \frac{T_0}{10^4 \text{ K}}\right)^{-0.72} \left(\frac{1+z}{7}\right)^{9/2} $\citep[see][for  a review]{Becker2015}, where $\Delta_b$ is the baryon overdensity and the temperature $T_0$ is assumed to be $10^4$ K.

At a constant mean density of $\Delta_b = 1$, the photoionisation rate due to A370p\_z1 is sufficient to have an average Lyman-$\alpha$ transmission in the bubble $\overline{\mathcal{T}_\alpha}(\rm{blue\,wing}) = 0.25(0.69)$, and a transmission at the edge of the blue peak $\mathcal{T}_\alpha^{\rm{blue\,peak}} = 0.51(0.93)$. We note this does not take into account expansion in the Hubble flow, ignores the effect of the IGM damping wing or overdensities associated with the galaxy, enhancements we consider beyond the scope of this discovery paper.
Recently, \citet{Mason2020} have laid out an extensive framework to model high-redshift double-peaked Lyman-$\alpha$ emitters. Their modelling suggests that a source with the luminosity of A370p\_z1 and $\fesc=1$ could carve an ionised bubble with $r_{\rm ion}\sim 0.6$ pMpc sufficient to permit the blue peak to escape up to $\sim 0.2$ pMpc, in good agreement with our results.

Finally, we checked that A370p\_z1 does not lie in an overdensity of $z\sim 6.8$ objects. We find 28 F105W-F814W dropout galaxies in the A370p field with $M_{\rm{ F125W}}<28$ whose $1\sigma$ photometric redshift is at least partially in the redshift interval $6.3<z<7.3$. This is in good agreement to that expected ($28\pm 5$) from the \citet{Bouwens2015} UV luminosity function. We thus conclude that A370p\_z1 is very likely to have contributed to the totality or the large majority of the LyC photons in its surrounding ionised bubble. 

\section{Discussion}

\subsection{Differences and similarities between NEPLA4, COLA1 and A370p\_z1}

We now apply the methodology described in the previous section to determine which $z>6$ double-peaks (A370p\_z1, COLA1 and NEPLA4) can grow an \hii\, bubble and ionise it sufficiently to permit blue peak photons to escape. We leave Aerith B aside as the ionised bubble created by the nearby quasar was studied in detail by \citet{Bosman2020}. For this exercise, we assume that the redshift of all three objects is taken from the mid-point of the two Lyman-$\alpha$ peaks. To further facilitate the comparison between objects, we assume an age of $10$ Myr for each galaxy (which matches the estimate for COLA1 in \citet{Matthee2018}, but is lower than what we measure for A370p\_z1). This only affects the Str\"omgren bubble radii which are proportional to $\propto t_{em}^{1/3}$ and can be rescaled accordingly if needed. The escape fractions, UV magnitudes and extent of the blue wings $r_\alpha$ presented in Table \ref{table:bubbles_doublepeaks} are taken from \citet[][]{Matthee2018,Songaila2018}, and private communication from A. Songaila.

Interestingly, we find that all double-peaks can grow a Str\"omgren sphere as large as the minimum bubble size $r_\alpha$ derived from the blue wing maximum velocity offset. However, only A370p\_z1 can grow a bubble that is $2-4$ times larger (depending on the ionising efficiency). This is important because the calculated radius of the Str\"omgren sphere is significantly larger than the maximum distance at which blue  photons still escape \citep{Mason2020}. The most significant test of whether a galaxy is self-ionising its local bubble, or if additional faint sources are needed to let the blue wing photons escape, is to compute the opacity to Lyman-$\alpha$ photons. We find that COLA1 and NEPLA4 are unable to solely ionised the CGM/IGM sufficiently to allow blue peak photons to escape. The predicted opacity at the  blue peak is $0.6\%$ in the most favourable scenario for COLA1, and always zero for NEPLA4. However, the blue peaks are clearly detected, with an observed blue/red peak flux ratio of 0.31 for COLA1 and 0.6 for NEPLA4. We conclude that additional sources are needed to maintain their \hii\, bubbles highly ionised. In contrast, A370p\_z1 is able to maintain its bubble sufficiently ionised on its own in all scenarios within the large 1$\sigma$ error of the peak flux ratio. Collectively, the four currently known high-redshifts double-peaks present a large range of cases from a source not contributing to reionisation (Aerith B) to a powerful source ionising its \hii\, bubble (A370p\_z1), and intermediate cases with significant $\fesc$ but probably surrounded with faint leakers which keep their \hii\, bubble highly ionised (NEPLA4,COLA1). 

\begin{table}
    \centering    
    \caption{Comparison of the ionising properties of the three known $z\sim6.5$ double-peaks. Str\"omgren radii are computed assuming ages of $10$ Myr.}
    \begin{tabular}{l|ccc} 
    & NEPLA4 & COLA1 & A370p\_z1 \\ \hline
    $M_{\rm{UV}}$ & $-21.8$ & $-21.6$ & $-21.5$ \\
    $\fesc(\Delta v)$ & $0.11$ & $0.29$ & $\approx 0.9$ \\
    EW$_{Ly\alpha}$ [\AA] & $176$ & $120$ & $43$  \\
    $r_\alpha$ [pMpc] & $0.31$ & $0.31$ & $0.26$ \\ \hline
    \multicolumn{4}{c}{Assuming $\xiion=10^{25.4}$ cgs} \\ \hline
    $r_S$ [pMpc]   & $0.29$ & $0.37$ & $0.50$ \\
    $\langle T_\alpha\rangle (\rm{blue\, wing})$ & $0.09$ & $0.13$ & $0.25$ \\
    $T_\alpha(\rm{blue\, peak})$ &  $1\times10^{-16}$  & $7\times10^{-5}$ &  $0.51$ \\ \hline
    \multicolumn{4}{c}{Deriving $\xiion$ from Lyman-$\alpha$  \citep{Sobral2019}}\\ \hline
    $\xiion$ &$10^{25.38}$ & $10^{25.66}$ & $10^{26.38}$ \\
    $r_S$ [pMpc]   & $0.28$ & $0.45$ & $1.05$ \\
    $\langle T_\alpha\rangle (\rm{blue\, wing})$ & $0.09$ & $0.17$ & $0.69$ \\
    $T_\alpha(\rm{blue\, peak})$ &  $3\times10^{-17}$  & $6\times10^{-3}$ &  $0.93$ \\ \hline
    Peak flux ratio & $\approx 0.6$ & $0.31\pm0.03$ & $0.93\pm0.28$
    \end{tabular}
    \label{table:bubbles_doublepeaks}
\end{table}

\subsection{Implications for reionisation}
\label{sec:conclusions}

We have shown that A370p\_z1 is possibly the first convincing example of a source capable, \textit{on its own}, of creating a significant ionised bubble \textit{and} maintaining this state so that photons escape bluewards of Lyman-$\alpha$. A key question, therefore, is whether it is an exceptional source or representative of a larger population of luminous objects responsible for concluding cosmic reionisation. Although many luminous $z>6$ galaxies have now been spectroscopically confirmed using the Lyman-$\alpha$ line \citep[e.g.][]{Zitrin2015,Laporte_2017a, Laporte_2017b,Stark2017,Songaila2018,Stark2018,Hashimoto2018a, Taylor2020}, the majority did not have the spectral resolution to resolve closely separated peaks as is the case in A370p\_z1. Additionally, very shallow blue peaks below the sensitivity limit of the observations would also be missed. Nonetheless, the unusually high confirmation rate of Lyman-$\alpha$ emission in the \citet{Roberts-Borsani2016} galaxies with strong IRAC 4.5$\mu$m excesses might be explained if they were efficient leakers that carved their own ionised bubbles \citep{Zitrin2015,Stark2017}. Alternatively, of course, there may be associated faint sources and/or AGN activity that contribute to the ionising flux.

Searching for a  larger sample of $z>6$ double-peaked Lyman-$\alpha$ emitters is therefore a promising way of studying both the sources of reionisation and their surrounding \hii\, bubbles with the growing modelling capabilities highlighted above. The rest-frame optical lines of these luminous $z>6$ sources will be detectable with \textit{JWST}, enabling us to characterise, amongst other quantities, their intrinsic ionising output.

\vspace{1cm}
\section*{Acknowledgements}
We wish to thank the anonymous referee for comments which significantly improved this manuscript. RAM, NL, RSE acknowledge funding from the European Research Council (ERC) under the European Union's Horizon 2020 research and innovation programme (grant agreement No 669253). NL also acknowledges support from the Kavli Foundation. AV acknowledges support from the SNF under the professorship grant 176808. AV and TG acknowledge support from the European Research Council under grant agreement ERC-stg-757258 (TRIPLE). RAM thanks K. Kakiichi, M. Gronke and J. Matthee for useful discussions. We are grateful to A. Songaila for sharing details about the blue wing of NEPLA4 and its Y band magnitude.\\
Based on observations made with ESO Telescopes at the La Silla Paranal Observatory under programme 0100.A-0664(A).
\section*{Data availability}
The data underlying this article are available in the ESO archive (archive.eso.org) under programme ID 0100.A-0664(A). 




\bibliographystyle{mnras}
\bibliography{ref} 




\appendix

\section{Non detections of rest-frame UV lines}
\begin{figure*}
    \centering
    \includegraphics[width = 0.49\textwidth]{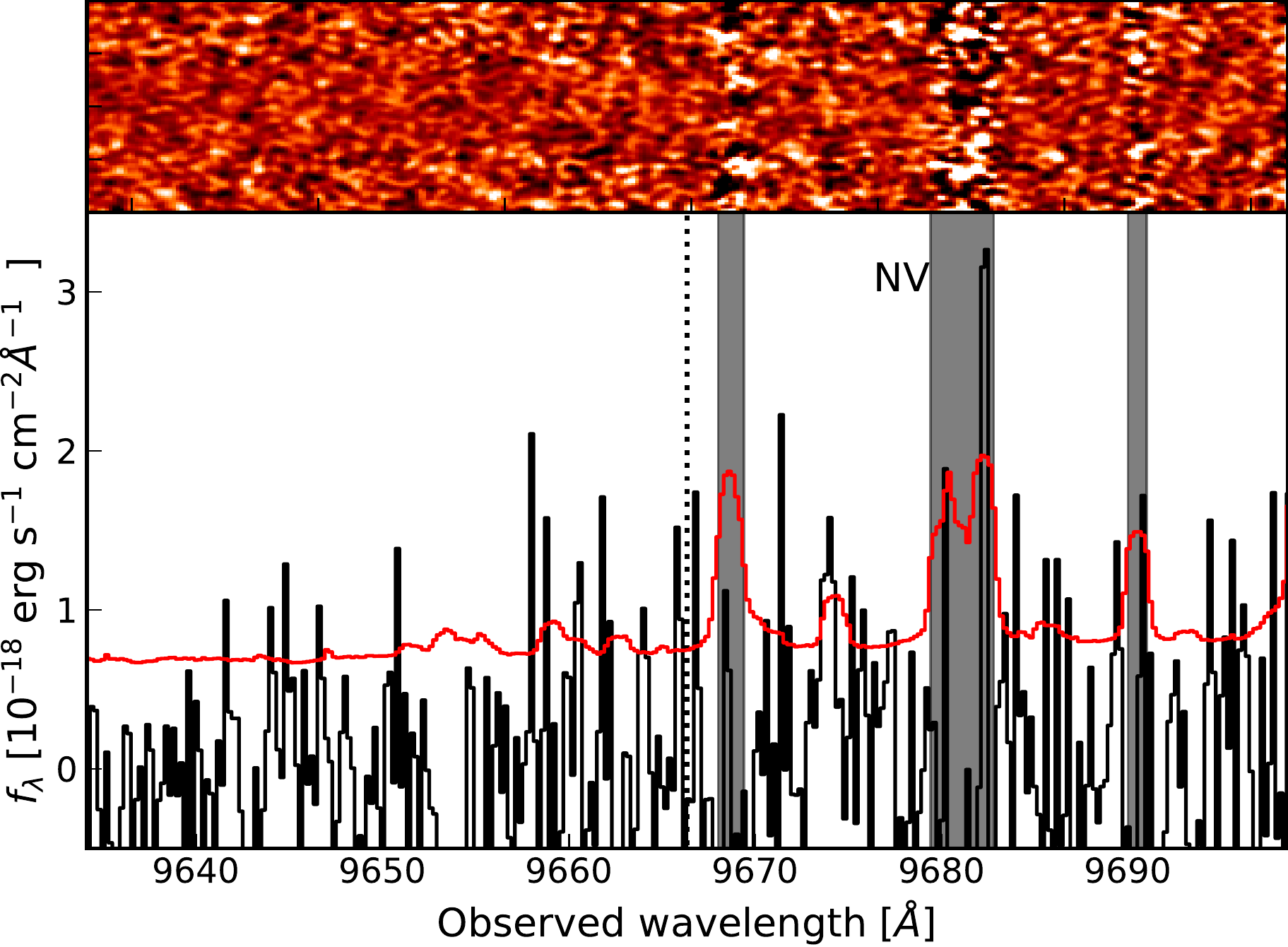} 
    \includegraphics[width = 0.49\textwidth]{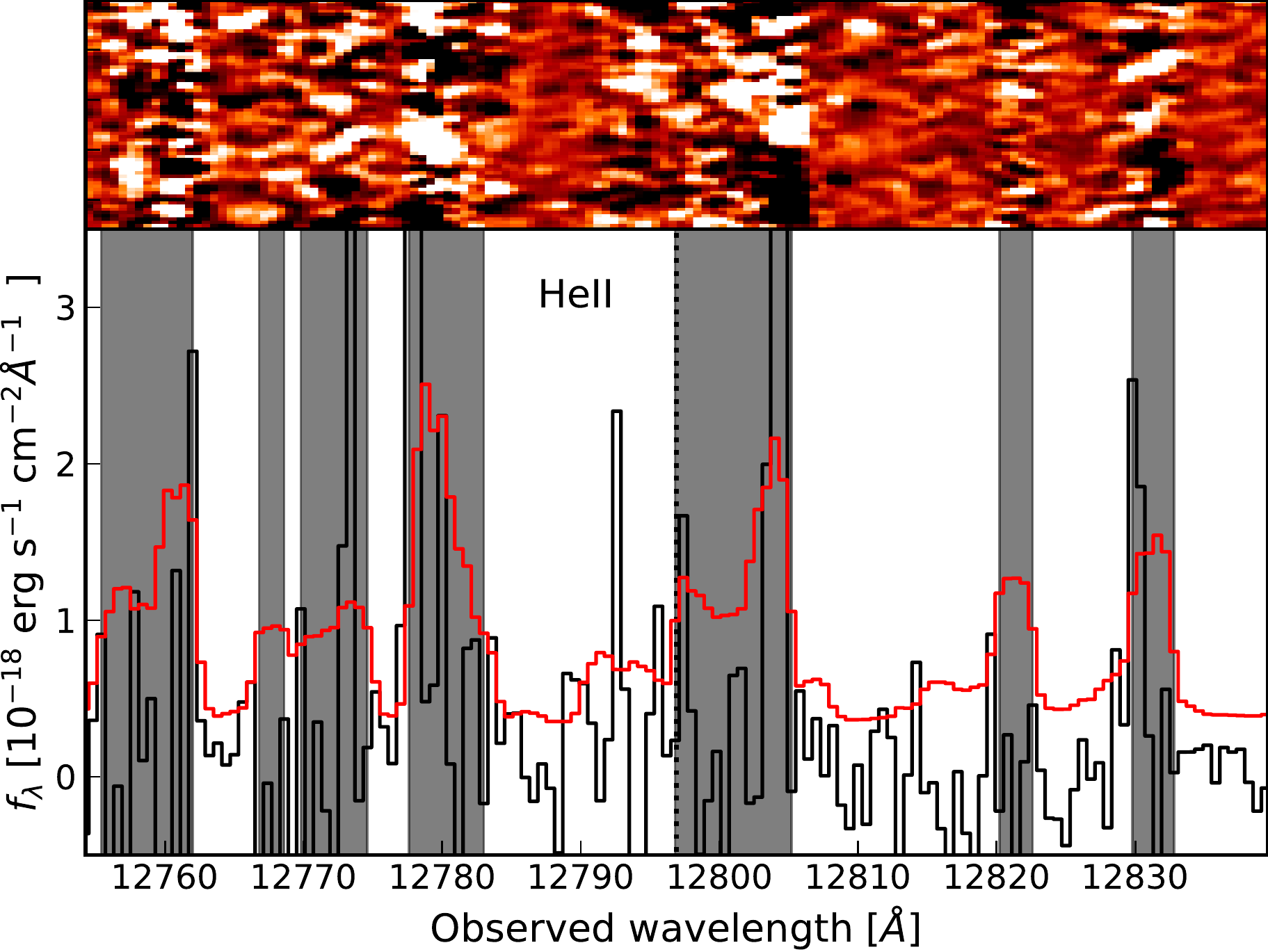} \\
    \includegraphics[width = 0.49\textwidth]{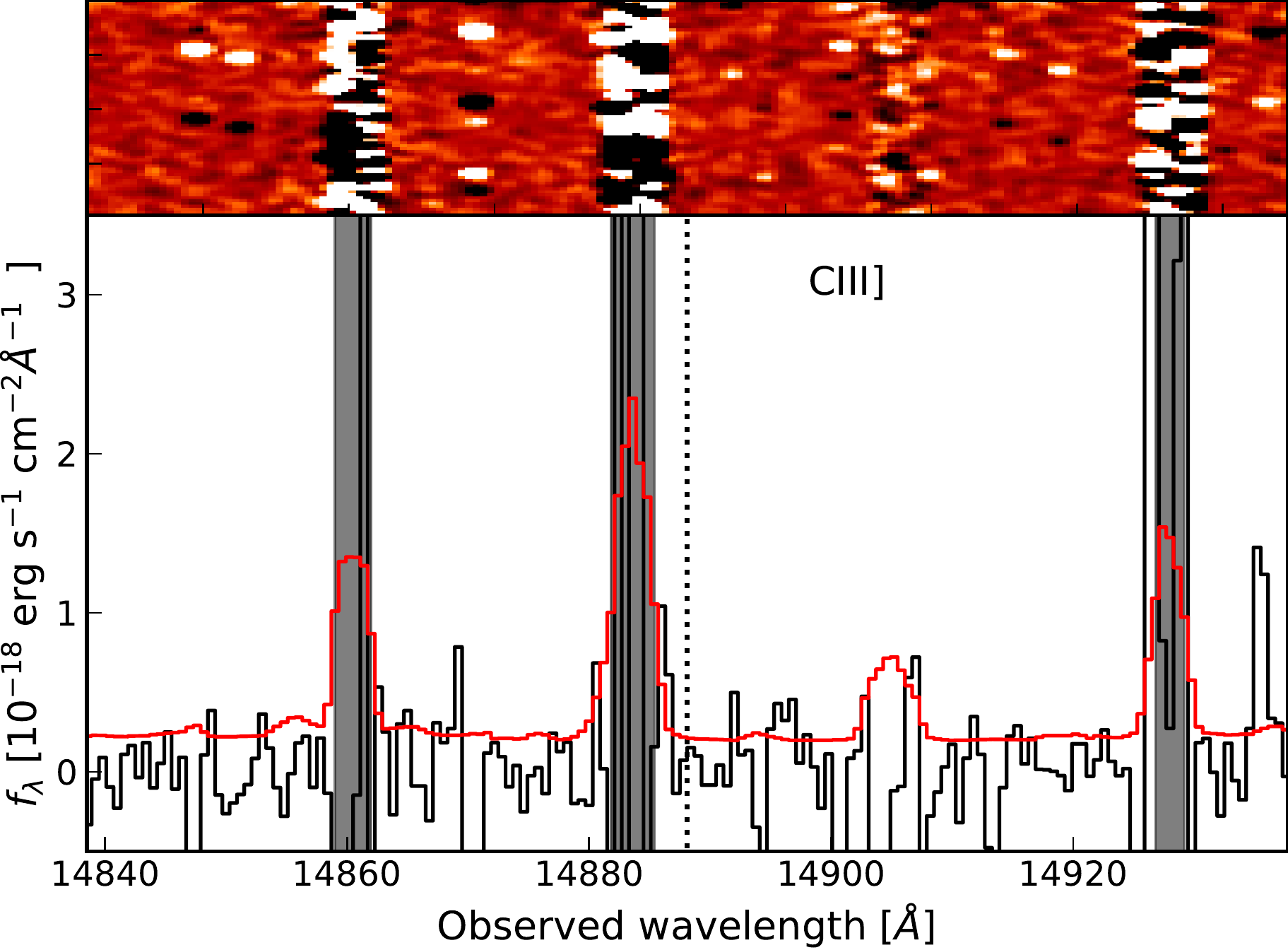}
    \includegraphics[width = 0.49\textwidth]{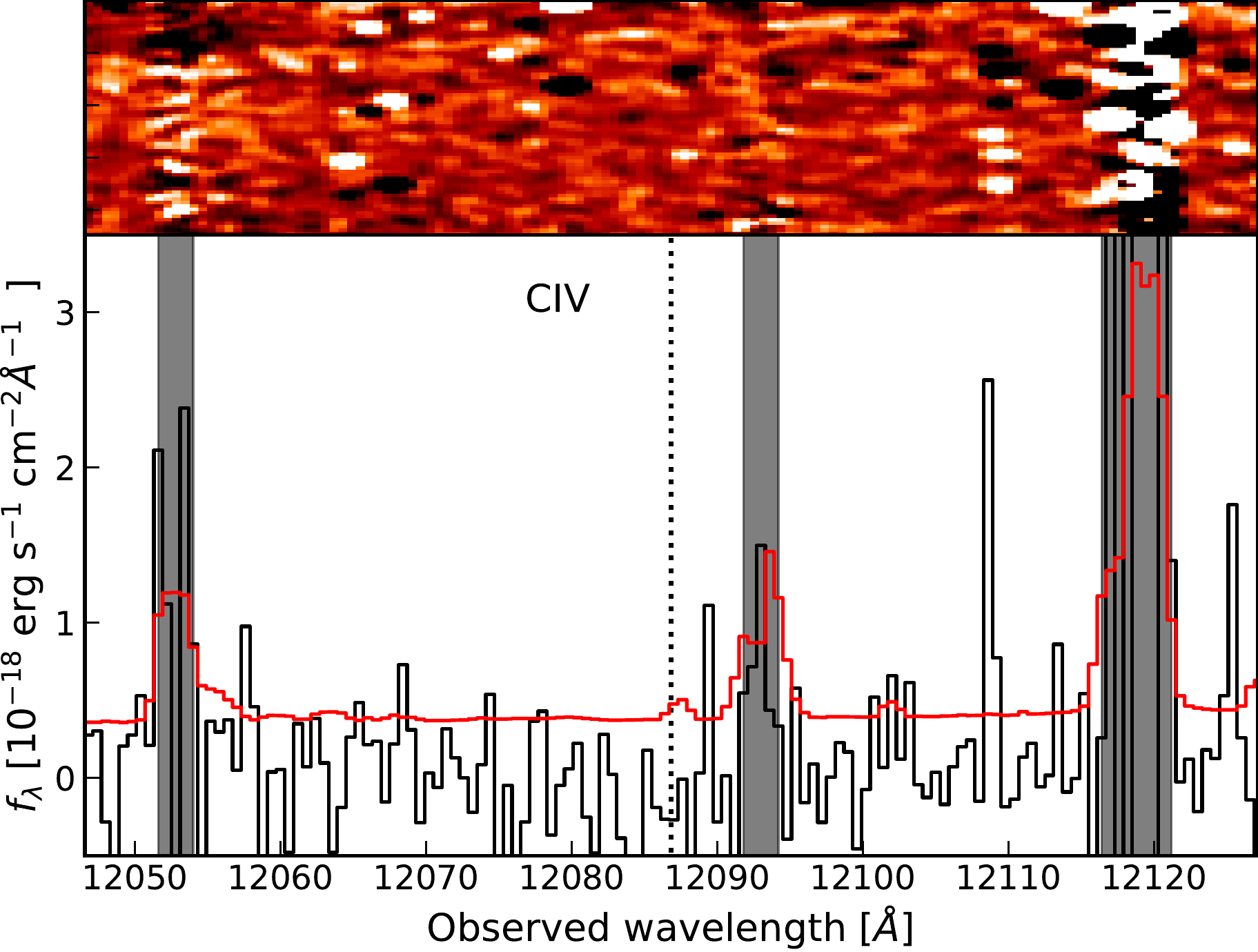}
    \caption{2D and 1D spectrum of A370p\_z1, showing the expected location of rest-frame UV lines at $z=6.803$ (x-axis range: $\pm 1000\,\kms$). Real emission lines should show one bright line at the center of the 2D spectra (upper panel) with two negative counterparts (black) at the top and bottom of the spectrum due to the ABBA nodding pattern adopted. The colour scheme is identical to that of Figure \ref{fig:observed}, but the smoothing length is adjusted for the NIR arm. The lower panel shows the 1D spectrum (black) and error array (red) with sky lines masked in grey. Vertical dotted lines show the exact wavelength of the UV lines or the centroid for doublets.  }
    \label{fig:observed_extralines}
\end{figure*}

We show on Figure \ref{fig:observed_extralines} the 1D and 2D spectrum of A370p\_z1 at the expected location of rest-frame UV lines. We do not find any significant lines (see Table \ref{table:properties} for detection limits)

\bsp	
\label{lastpage}
\end{document}